\documentclass[a4,aps,twocolumn,showpacs,preprintnumbers,qamslatex,amsmath,amssymb,longbibliography]{revtex4-1}
\usepackage[cp1251]{inputenc}
\usepackage[T2A]{fontenc}
\usepackage{fancyhdr}
\usepackage[colorlinks]{hyperref}
\usepackage{graphicx}
\usepackage{verbatim}
\usepackage{indentfirst}
\usepackage{latexsym}
\usepackage{dsfont}
\usepackage{epstopdf}

\begin{document}

\title{Dimerization in ultracold spinor gases with Zeeman splitting}
\author{Michael Weyrauch$^1$ and Mykhailo V. Rakov$^2$}
\affiliation{$^1$ Physikalisch-Technische Bundesanstalt, Bundesallee 100, D-38116 Braunschweig, Germany}
\affiliation{$^2$ Kyiv National Taras Shevchenko University, 64/13 Volodymyrska Street, Kyiv 01601, Ukraine}

\begin{abstract}
Two recent publications report different boundaries for the dimerized phase of the bilinear-biquadratic spin-1 Heisenberg model with quadratic Zeeman effect. We address these discrepancies for the  biquadratic model with quadratic Zeeman term and explain the differences. Based on our numerical results the phase boundaries of the
dimerized phase are determined.
\end{abstract}


\maketitle

\section{Introduction}

Ultracold atoms trapped in optical lattices are ideally suited for investigations of
the phase structure and phase transitions of strongly interacting quantum many-body systems.
A recently highlighted example is the superfluid to Mott insulator transition studied in Ref.~\cite{Greiner2002} using Rubidium Bose-Einstein condensates. In the Mott insulating phase the
spinor gases show a multitude of magnetic phases thus providing amble possibilities for studies of
quantum magnetism in different dimensionalities and insight into conventional as well as topological phases.

In certain limits such systems may be modelled by spin lattice Hamiltonians
$
H=\sum_{i=1}^N \, h_{i,i+1}
$
with nearest-neighbor interactions only~\cite{PhysRevLett.93.250405}.
A prominent example is the one-dimensional bilinear-biquadratic spin-1 Heisenberg model with quadratic Zeeman term
\begin{equation}\label{ham}
h_{i,i+1}=\cos\theta \, \vec S_i \otimes \vec S_{i+1} + \sin\theta \, (\vec S_i \otimes \vec S_{i+1})^2+D \, (S_i^z)^2
\end{equation}
and $S^\nu_i$ the spin-1 SU(2) matrix representations ($\nu=x,y,z$, and $ i=1,\ldots, N$ with $N+1 \rightarrow 1$). This model shows a rich phase structure, and a
rather complete overview  was
recently given by Rodriguez~{\it et al.}~\cite{PhysRevLett.106.105302} and De Chiara, Lewenstein, and Sanpera~\cite{PhysRevB.84.054451} as a function of $\theta$ and the Zeeman strength $D$.

Despite qualitative agreement, the results of the two groups disagree significantly concerning the extension of the dimerized phase. In Ref.~\cite{PhysRevB.84.054451} the dimerized phase extends from some undefined $D < -2$ up to about $D\simeq 0.03$ at $\theta=-\pi/2$. On the contrary, the authors of Ref.~\cite{PhysRevLett.106.105302} find a dimerized phase in the parameter range $-0.3 \lesssim D \lesssim 0.6$ at this $\theta$. The methods employed in both papers are rather different: In Ref.~\cite{PhysRevLett.106.105302} the boundaries of the dimerized phase are obtained using level spectroscopy~\cite{Okamoto1992} in relatively small spin rings ($N\leq16$), while
in Ref.~\cite{PhysRevB.84.054451} the dimerization order parameter is calculated from numerically obtained ground states of spin chains up to $N=204$.

It is the purpose of the present paper to address this discrepancy using variants of both methods in parallel.
To this end we determine both the spectrum as well as the order parameter at $\theta=-\frac{\pi}{2}$ as a function of the Zeeman coupling $D$. Calculations will be performed for systems with periodic boundary conditions (spin rings) using our own matrix product state (MPS) algorithm for systems up to 100 sites~\cite{Weyrauch2013, PhysRevB.93.054417, Rakov2017}.

At $\theta=-\frac{\pi}{2}$ and $D=0$ only the biquadratic term remains in (\ref{ham}) which is  SU(3) symmetric~\cite{Affleck1986,PhysRevB.65.180402}, i.e. it may be rewritten as a bilinear model in terms of the three-dimensional Gell-Mann SU(3) `quark' ($\lambda$) and `antiquark' ($\bar{\lambda}$) triplet representations,
\begin{equation}\label{ham1}
h_{i,i+1}=-8  \vec{\lambda}_i\otimes\vec{\bar{\lambda}}_{i+1}-\frac{4}{3} \, \mathds{1}.
\end{equation}
The quadratic Zeeman term, which in terms of Gell-Mann matrices reads as $2D (1/3 \, \mathds{1} + \lambda_3 - \lambda_8$), reduces the symmetry from SU(3) to SU(2), i.e. the Gell-Mann triplet splits into an SU(2) spin-$\frac{1}{2}$ duplet and one singlet at each site. We shall call this SU(2) subgroup $v$-spin. This SU(2) symmetry holds only at $\theta=-\pi/2, \pi/2, -3\pi/4$, and  $\pi/4$, and is different from the $D=0$ SU(2) symmetry of the Hamiltonian~(\ref{ham}), which we shall call $s$-spin. The latter reduces to U(1) at all $\theta$ due to the Zeeman term.

Since a continuous symmetry cannot be broken in one dimension~\cite{PhysRevLett.17.1133,Coleman1973}, we developed a matrix-product algorithm which incorporates $v$-spin symmetry explicitly in the ansatz for the MPS similar to our treatment of SU(2) symmetric MPS presented in Ref.~\cite{Rakov2017}. As a consequence, the obtained states may be labeled by SU(2) $v$-spin quantum numbers as will be done in this paper. The U(1) subgroups of $v$-spin and $s$-spin are related by $S_z=2 v_z$.

In the following section we study the low-lying spectrum at $\theta=-\pi/2$ for various system sizes $N$ and anisotropy parameters $D$ and extrapolate these results to the thermodynamic limit. The extension of the dimerized phase is then determined from the parameter region where the extrapolated ground state energy is doubly degenerate. In addition,
we also calculate the dimerization order parameter in this parameter region and compare both results for consistency. We also determine the nematic order parameter in this phase.

It is well known that for $D=0$ the bilinear-biquadratic spin-1 system is dimerized for all $\theta$ between the two critical points  $\theta=-\frac{3\pi}{4}$ and $\theta= -\frac{\pi}{4}$.
Using the results at $\theta=-\pi/2$ as a guidance, we phenomenologically extrapolate our results to this parameter region. This extrapolation is summarized in Fig. 5 of the present paper, and it will be discussed
in detail in the summary section.

\section{Boundaries of the dimerized phase of the biquadratic Heisenberg model with quadratic Zeeman term}~\label{determination}

The boundaries of the
dimerized phase have been studied by Rodriguez {\it et al.}~\cite{PhysRevLett.106.105302} using level spectroscopy and
by De Chiara {\it et al.}~\cite{PhysRevB.84.054451} from a direct calculation of the dimerization order parameter. The results are surprisingly different.

The spectra for large enough systems indicate phase boundaries by the closing or opening of spectral gaps. Since
we only determine spectra for finite systems, we find level crossings which may or may not
indicate the closing or opening of spectral gaps in the thermodynamic limit.

\begin{figure}
\unitlength 1cm
\includegraphics[width=0.4\textwidth]{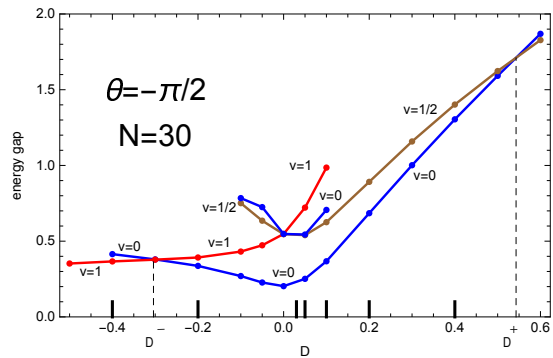}
\caption{\footnotesize Low lying spectrum of the biquadratic ($\theta=-\pi/2$) Heisenberg ring with $N=30$ spins and quadratic Zeeman term in the parameter range $-0.5 < D < 0.6$. The two lowest $v=0$ excitations and the lowest $v=1/2$ and $v=1$ multiplets are shown (the ground state is a singlet shifted to $E=0$). At $D=0$, one observes one SU(3) octet above the two low lying SU(3) singlets, which splits into SU(2) $v$-spin multiplets at $D\neq0$. There are two characteristic level crossings at $D^- \simeq -0.30$ and $D^+ \simeq 0.54$ indicated by dashed black vertical lines. The long thick tick marks along the horizontal axis indicate those values of $D$ at which we calculate spectra for larger systems. The essential structure of the spectrum remains very similar for larger systems due to $v$-spin symmetry.
\label{spectrum30}}
\end{figure}

The spectrum for $N=30$ sites and $\theta=-\pi/2$ is shown in Fig.~\ref{spectrum30} as a function of $D$ in the interval $-0.5 < D < 0.6$. Characteristic level crossings are indicated in Fig.~\ref{spectrum30} by the dashed black lines at $D=D^-$ and $D=D^+$. These lines agree rather well with the dimerized phase boundaries $D^-$ and $D^+$ obtained in Ref.~\cite{PhysRevLett.106.105302} for this $\theta$. (Note, that due to a different sign convention for the Zeeman term "$+$" and "$-$" must be interchanged when comparing to our results.)
The parameter region $D<D^-$ is characterized as the boundary of a XY nematic phase~\cite{PhysRevLett.106.105302} and its lowest excitation is a $v$-spin triplet (see Fig.~\ref{spectrum30}). The region $D>D^+$ is characterized as an Ising nematic phase, and its lowest excited states are two degenerate $v$-spin doublets. For $D^-<D<D^+$ the lowest excited state is a singlet, and in the thermodynamic limit one expects dimerization if the gap between the
two lowest singlets closes. In the following we will investigate in detail, if this scenario suggested in Ref.~\cite{PhysRevLett.106.105302} sustains detailed scrutiny.

\subsection{Low lying spectrum}

\begin{figure}
\unitlength 1cm
\includegraphics[width=0.4\textwidth]{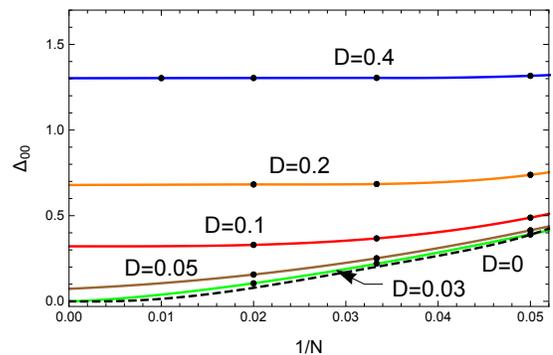}
\caption{\footnotesize  Energy gap between the two lowest $v=0$  states of the biquadratic Heisenberg ring with quadratic Zeeman term for various positive $D$. The extrapolated gaps $\Delta_{00}^{\infty}$ are finite for $D \ge0.03$ (e.g., $\Delta_{00}^{\infty} (D=0.05) \simeq 0.07$).
\label{gap}}
\end{figure}

\begin{figure*}
\unitlength 1cm
\begin{picture}(15,6)(0,0)
 \put(0,0)    {\includegraphics[width=0.4\textwidth]{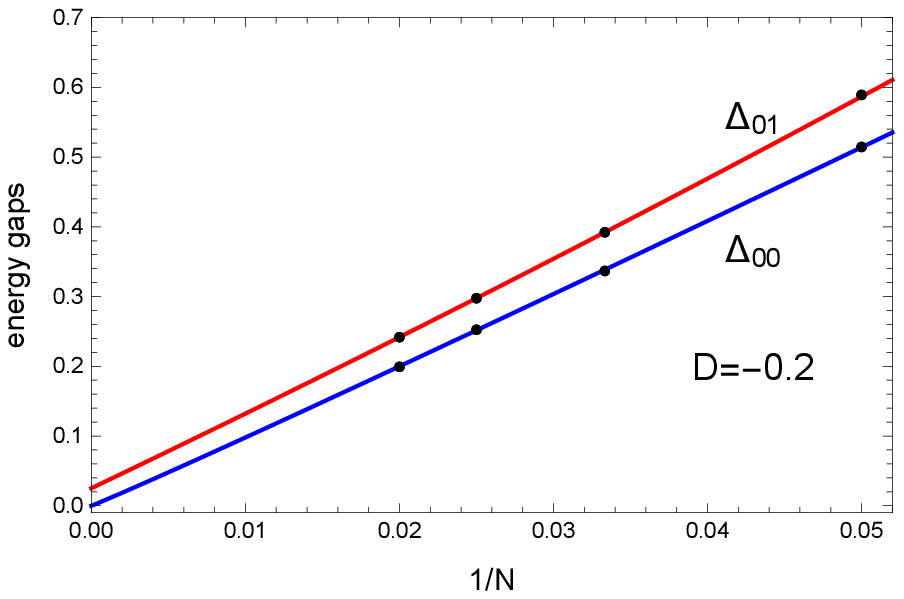}}
 \put(7.5,0)  {\includegraphics[width=0.4\textwidth]{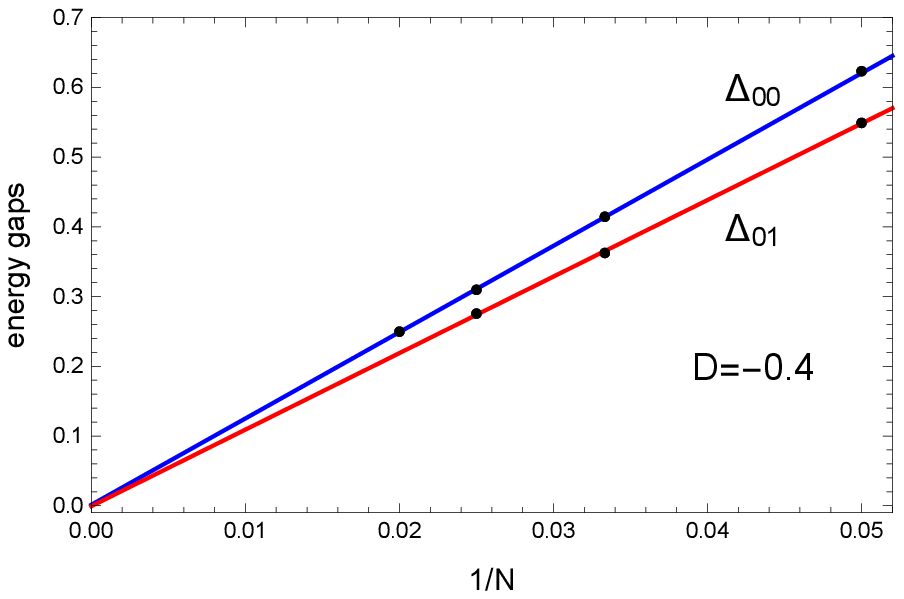}}
\end{picture}
\caption{\footnotesize Energy gap between the two lowest $v=0$ states and between the lowest $v=0$ state and the $v=1$ multiplet of the biquadratic Heisenberg ring with quadratic Zeeman effect at $D =-0.2$ (left) and $D=-0.4$ (right). The extrapolated gap $\Delta_{00}^{\infty} \simeq 0$ for both values of $D$, which indicates the presence of the dimerization. The extrapolated gap $\Delta_{01}^{\infty} (D=-0.2) \simeq 0.025$ is finite, while both gaps are closing in the thermodynamic limit at $D=-0.4$. The small value of $\Delta_{01}^{\infty} (D=-0.2)$ is in line with the suggestion that the phase transition at $D^-$ is of Kosterlitz-Thouless type.
\label{gap-02-04}}
\end{figure*}

We first study the low lying spectrum as a function of system size $N$ at several characteristic $D$ indicated by the
large tick marks along the horizontal axis in Fig.~\ref{spectrum30}. Our results are collected in
Figs.~\ref{gap}-\ref{gap-02-04}. We consider system sizes between $N=20$ up to $N=100$.

The finite size dependence of the spectrum at $D=0$ was studied extensively by S{\o}rensen and Young~\cite{PhysRevB.42.754} using the Bethe Ansatz. In the thermodynamic limit the gap $\Delta_{00}$ between the lowest two SU(2)/SU(3) singlets closes while the gap to the SU(3) octet ($v$-spin triplet) remains finite. We include these $D=0$ results  in Fig.~\ref{gap} for comparison (dashed black line).


From the results for  shown in Fig.~\ref{gap} we conclude that for positive $D \ge 0.03$ the gap does not close in the thermodynamic limit. As a consequence, the system does not dimerize due to translational symmetry.
This result agrees with the findings of De Chiara {\it et al.}~\cite{PhysRevB.84.054451} from the calculated dimerization in finite chains as will be discussed in more detail in the next subsection. The question if the gap $\Delta_{00}$ closes for $0<D \le 0.03$ cannot be decided by our numerics. However, from the results presented in
in the following subsection we expect that the gap closes in the region $0<D \lesssim 0.025$.

We now consider the parameter range $D<0$. It is worth mentioning that much larger computational resources are required here than for positive $D$, since the correlation length is larger. In practice we gradually increase the degeneracy set until the result is converged. In the region $D \lesssim -0.3$ the correlation length increases monotonically with the system size, and the results are numerically rather hard to obtain.

In Fig.~\ref{gap-02-04}~(left) we show the energy gaps $\Delta_{00}$ and $\Delta_{01}$ for $D=-0.2$ as a function of $1/N$. The extrapolated results indicate that $\Delta_{00}$ closes and $\Delta_{01}$ remains open in the thermodynamic limit. This is similar to the facts found in~\cite{PhysRevB.42.754} at $D=0$, and it indicates that the point $D=-0.2$ is inside the dimerized phase.

The results for $D=-0.4$  shown in Fig.~\ref{gap-02-04}~(right) suggest that {\it both} gaps $\Delta_{00}$ and $\Delta_{01}$ are closing in the thermodymanic limit. Consequently, the system is still dimerized at $D=-0.4$ with additional gapless nematic excitations. In fact, our results suggest, that a Kosterlitz-Thouless transition to a critical nematic phase happens exactly at $D^-$. However, the phases on both sides of this transition are dimerized. 
The dimerization does not signal this transition.

\subsection{Dimerization and nematics}

De Chiara {\it et al.}~\cite{PhysRevB.84.054451} determined the extension of the dimerized phase from the expectation value of the dimerization operator
$
\hat{\mathcal{D}}=\frac{1}{N} \, \sum_i \, (-1)^i \, h_{i,i+1}
$
calculated for finite chains up to $N=204$ and extrapolated to the thermodynamic limit.

For finite  rings the ground and the excited states {\it cannot} be dimerized due to translational invariance. In order to calculate the dimerization, a symmetric/antisymmetric superposition  $\frac{1}{\sqrt{2}}| 0^{(0)} \pm 0^{(\pi)}\rangle$ of the two lowest $v=0$ states with different momenta ($p=0,\pi$) is taken. These two states are separated by a small gap for finite systems, but they develop into a degenerate doublet in the thermodynamic limit within the dimerized phase. It is important to make sure that the lowest two $v=0$ states are {\it indeed} degenerate in the thermodynamic limit before using this procedure.

\begin{figure*}
\unitlength 1cm
\begin{picture}(15,6)(0,0)
 \put(0,0)    {\includegraphics[width=0.4\textwidth]{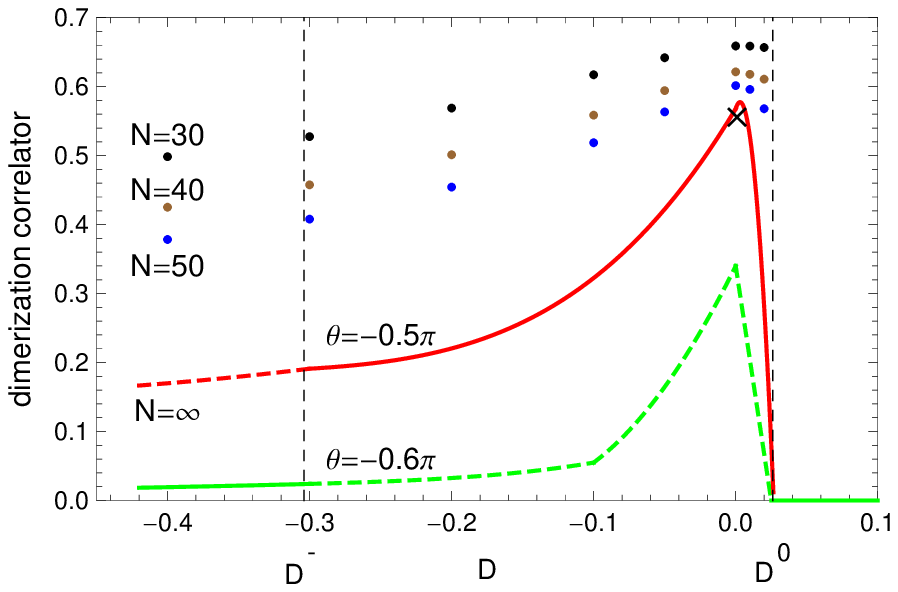}}
 \put(7.5,0)  {\includegraphics[width=0.4\textwidth]{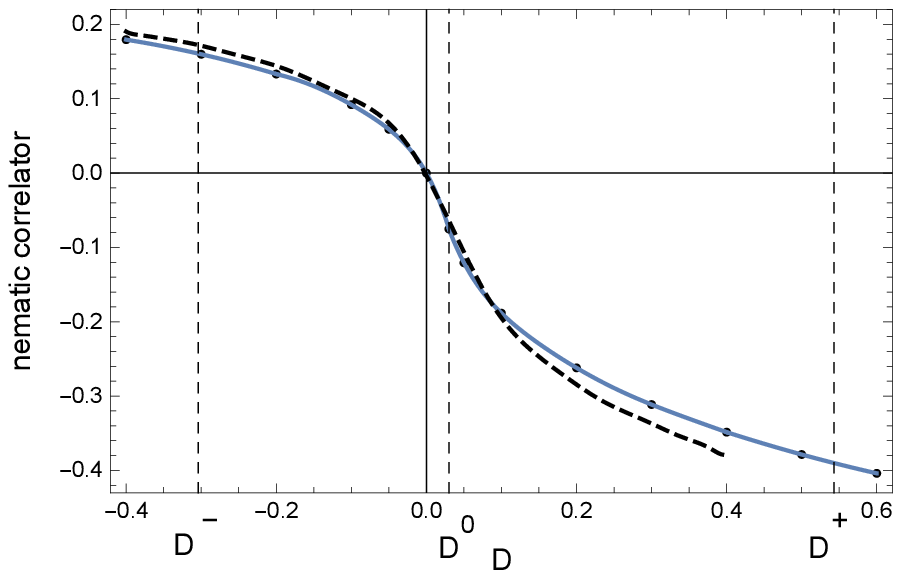}}
\end{picture}
\caption{\footnotesize  (Left) Dimerization correlator calculated for the biquadratic Heisenberg model with quadratic Zeeman effect from the two lowest $v=0$ eigenstates with different momenta. Results for 30, 40, and 50 sites are presented. The red line shows the extrapolation to the thermodynamic limit.  The black cross indicates the Bethe ansatz value at $D=0$ and $N \rightarrow \infty$. The dimerization is strictly zero at $D \gtrsim 0.025$. The green line shows the result obtained by De Chiara {\it et al.}~\cite{PhysRevB.84.054451} at $\theta=-0.6 \pi$ (the dashed part is extrapolated from their data). Our results  confirm the prediction of De Chiara {\it et al.} that the dimerization persists up to large negative values of $D$. (Right) Nematic correlator extrapolated to the thermodynamic limit (blue line). It is calculated from the two lowest eigenstates for $-0.3 \lesssim D \lesssim 0.025$ and from the five lowest eigenstates for $D \lesssim -0.3$. The nematic correlator is exactly zero at $D=0$. The inflection point $D \simeq 0.02$ coincides rather well with the dimer-to-non-dimer phase transition point $D^0 \simeq 0.03$. The nematic correlator is featureless at $D^-$ and $D^+$. The results obtained by Rodriguez {\it et al.}~\cite{PhysRevLett.106.105302} at $\theta=-0.54\pi$ for spin chains of 36 sites are included for comparison (black dashed line).
\label{dimernematic}}
\end{figure*}

Our results for the dimerization correlator are presented in Fig.~\ref{dimernematic}.
For $D=0$ the dimerization is well-known from the literature~\cite{Xian1993, Baxter1973},  $\mathcal{D}_{\infty}=\frac{\sqrt{5}}{2} \prod_{n=1}^\infty \tanh^2(n \, {\rm arccosh} \frac{3}{2}) \simeq 0.562$. Our results for 30, 40 and 50 sites at $D=0$ can be fitted very well by the function~\cite{PhysRevB.42.754, PhysRevB.72.054433}
$
\mathcal{D}(N)=\mathcal{D}_{\infty}+c\,N^{-\alpha}\,\exp(-N/2\xi)
$
with $\alpha=1$. From the fit we obtain $\mathcal{D}_{\infty} \simeq 0.568$ and a large  correlation length $\xi \simeq 20.2$ in good agreement with the Bethe Ansatz. A very similar result for the correlation length was obtained in Ref.~\cite{PhysRevB.42.754} from the lowest energy gap. In order to confirm that the system dimerizes for small positive  $D$
we made detailed calculations for $D=0.01$ and $D=0.02$, where level spectroscopy was inconclusive, and clearly find nonvanishing dimerization.
The green line in Fig.~\ref{dimernematic} shows results by De Chiara {\it et al.} at $\theta=-0.6\pi$ (somewhat extrapolated by us).
In general, we confirm the result of De Chiara {\it et al.} that the dimerization extends from $-\infty<D \lesssim 0.025$.

In addition to the dimerization, we also present results for the nematic correlator (`chirality') $Q=\frac{1}{N} \, \sum_i (S_i^z)^2 -\frac{2}{3}$ of the ground state. At $\theta=-\pi/2$ and $D=0$ the nematic correlator is zero for any system size~\cite{PhysRevLett.106.105302}. In order to extrapolate the results to the thermodynamic limit, one must take into account that the ground state in the thermodynamic limit is 2-fold degenerate at $-0.3 \lesssim D \lesssim 0.025$ and  5-fold degenerate at $D \lesssim -0.3$. Unlike the dimerization correlator, the nematic correlator of any eigenstate is nonzero while the expectation value calculated from two different eigenstates is zero. It is observed that nematic correlators of each of the 2 (or 5) states are equal to a high precision for large systems, and finite-size effects are small. Therefore, a precise extrapolation to $N \rightarrow \infty$ is possible from calculations for systems of $N \le 50$ sites.

The nematic correlator shows a characteristic inflection point at $D \simeq 0.02$ (a similar inflection was obtained for $\theta=-0.54\pi$ in~\cite{PhysRevLett.106.105302}). This inflection appears to signal the transition from a dimerized to non-dimerized phase. On the other hand, the nematic correlator is featureless at $D^-$ and $D^+$.

Finally, we confirmed that the staggered magnetization of the ground state is zero throughout the line $\theta=-\pi/2$. This is not in contradiction with our suggestion that at $D<D^-$ the line $\theta=-\pi/2$ is a critical-to-Neel phase boundary.

\section{Conclusions}\label{conclusions}

\begin{figure}[t!]
\unitlength 1cm
\includegraphics[width=0.45\textwidth]{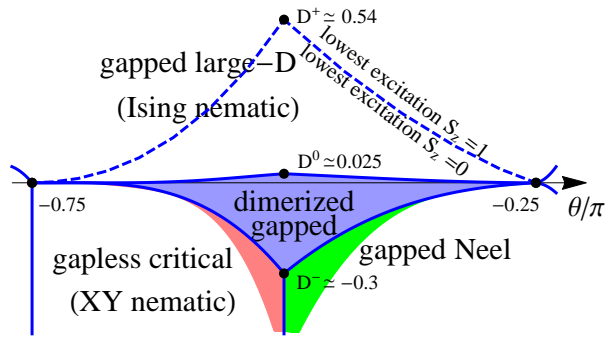}
\caption{\footnotesize Schematic phase diagram of the bilinear-biquadratic Heisenberg model with quadratic Zeeman anisotropy $D$ in the range $-\frac{3\pi}{4} < \theta < -\frac{\pi}{4}$. The three coloured phases (red, green and blue) are dimerized, full blue lines indicate phase transitions. Details are discussed in the main text.}\label{phased}
%

\end{figure}

In this work we numerically obtained  the boundaries of the dimerized phase of the biquadratic ($\theta=-\frac{\pi}{2}$) spin-1 Heisenberg model with quadratic Zeeman anisotropy. We find that a {\it gapped} dimerized phase exists in the parameter range $D^- < D < D^0$ with $D^-\simeq-0.30$ and $D^0\simeq 0.025$.
Moreover, we identify a $\it gapless$ dimerized phase which extends from $D^-$ to large negative $D$.  While this confirms the  results of De Chiara {\it et al.}~\cite{PhysRevB.84.054451}, who predicted a small dimerization even below $D<-2.0$, the existence of both a gapped and a gapless dimerized phase is reported here for the first time.   The transition between these two dimerized regions occurs at $D^-$ which was erroneously identified as the boundary of the dimerized phase in Ref.~\cite{PhysRevLett.106.105302}.
At the upper end of the dimerized region close to $D^0$, the dimerization sharply drops to zero and a gap opens between the two lowest singlet states marking the transition to a non-dimerized phase. We do not see a phase transition at $D^+$ which was identified as the upper dimerized phase boundary in Ref.~\cite{PhysRevLett.106.105302}. These findings for $\theta=-\frac{\pi}{2}$ are graphically represented on the vertical axis of the phase diagram shown in Fig.~\ref{phased}, where the various transition points are marked by black dots.

Let us now qualitatively extrapolate these results for $\theta$ in the parameter interval $I=[ -\frac{3\pi}{4} , -\frac{\pi}{4}]$,  separately for positive $D$ and negative $D$, guided by general considerations and the calculations presented in Refs.~\cite{PhysRevLett.106.105302} and \cite{PhysRevB.84.054451}: By now it is rather well established that for $D=0$ the bilinear-biquadratic spin-1 model has a dimerized ground state in the whole parameter interval $I$. In particular, a nematic non-dimerized phase close to the ferromagnetic transition has been ruled out~\cite{PhysRevLett.98.247202,PhysRevLett.113.027202}. At large negative or positive  $D$ the system is not dimerized. This follows from simple analytical arguments~\cite{PhysRevB.84.054451}.

At large positive $D\gg 1$ the system is in the gapped large-$D$ phase~\cite{PhysRevB.84.054451}, and the transition between the dimerized phase to a non-dimerized phase happens at small positive $D$ for all $\theta \in I$~\cite{PhysRevB.84.054451}. This we confirmed in the present paper for $\theta=-\pi/2$. In fact, one expects that the system is in an Ising nematic phase for all $D>D^0$ as indicated by the white region above the blue shaded region in Fig.~\ref{phased} as there are no gaps closing in the spectrum. However, it is expected  from the results of Ref.~\cite{PhysRevLett.106.105302} that the leading excitation changes from $S_z=0$ for $D<D^+$ to $S_z=\pm 1$ at $D>D^+$ as indicated by the blue dashed lines in the phase diagram.
According to Ref.~\cite{PhysRevLett.113.027202} the dimerization is related to the density of disclinations created in the spin system. Consequently, such topological defects should be absent for $D>D^0$.

For large negative $D$ the system is in a gapless critical (XY nematic) phase for $-\frac{3\pi}{4}<\theta<-\frac{\pi}{2}$ and in a gapped Neel phase for $-\frac{\pi}{2}<\theta<\frac{\pi}{2}$ ~\cite{PhysRevB.84.054451}. For small and intermediate negative $D$, the  XY nematic  and the Neel phases are separated by dimerized phases as  indicated by the blue, red, and green shaded regions in Fig.~\ref{phased}. It is expected that the gap between the lowest two singlets closes in all theses colored regions making them dimerized. In addition, in the red region also the gap to the next triplet closes, i.e. one expects a 5-fold degenerate ground state and vanishing staggered magnetization. This corresponds to our findings  at $\theta=-\frac{\pi}{2}$.
In the dimerized green shaded region, we expect an open gap to the triplet state and a non-zero staggered magnetization. The line between the red and green dimerized sectors separates magnetically staggered and non-staggered phases. This must be confirmed in detail by further calculations.
These considerations and extrapolations are summarized in the qualitative phase diagram shown in Fig.~\ref{phased}.

\begin{acknowledgments}
We thank Alexei K. Kolezhuk for discussions.
Mykhailo V. Rakov thanks Physikalisch-Technische Bundesanstalt for financial support during short visits to Braunschweig.
\end{acknowledgments}

\end{document}